\documentclass[sigconf,nonacm]{acmart}
\newcommand{\alextodo}[1]{}   %
\setcopyright{none}
\acmConference[ICAIF '26]{7th ACM International Conference on AI in Finance}{November 14--17, 2026}{Milan, Italy}
\begin{document}

\title{Velocity- and Regime-Aware Detection of Intraday Options Market
Manipulation, with Explainable Attribution}

\author{Alexander Chen}
\affiliation{\institution{University of Georgia}\city{Athens}\state{GA}\country{USA}}
\author{Maria Hybinette}
\authornote{Corresponding author.}
\affiliation{\institution{University of Georgia}\city{Athens}\state{GA}\country{USA}}

\begin{abstract}
Intraday market manipulation is hard to detect because its footprint is
brief, buried in millions of quotes, and statistically similar to ordinary
volatility. 
Detectors reach high recall only by flagging so many other days that measured 
precision collapses, producing alerts no regulator can act on.
We show that this manipulation leaves a distinctive \emph{dynamic
signature}: a pump-and-crash pattern visible in the
\emph{velocity} of market state, rather than its level.

We build a minute-level detection pipeline, strictly partitioned in
time, and based on smoothed state velocity: option-Delta velocity for
index options and price velocity for equities.
We explain every alert with SHAP attribution.
We hold the test period strictly out-of-sample and fix all thresholds
before evaluation.
On the locked Indian BANKNIFTY index-options test, the plain autoencoder
recovers 10 of 10 regulator-identified manipulation days.

Conditioning detection on market regimes inferred by a hidden Markov
model yields an instructive negative result. The regimes are descriptively distinct,
but using them trades recall for precision. Under the closed-world
assumption that all unlabeled days are normal, precision remains near
25\%.

The same dynamic appears in thinly traded U.S. equities
(SEC v.\ Patel, 2026). 
The \emph{shape} of the signature survives the
transfer; its velocity magnitude does not. 
A pump-reversal shape score ranks the complaint's alleged manipulation 
days with an area under the receiver operating characteristic curve (AUC) 
of 0.91 (ARQQ) and 0.81 (ACY).
On the ARQQ worked example, the score 
peaks inside the minute window documented in the complaint.

Finally, exact SHAP attribution over every alert shows that unconfirmed
alerts share the regulator-identified days' attribution profile (cosine
similarity 0.99).
The precision ceiling is consistent with incomplete
enforcement labels rather than detector failure.
What transfers across markets and instrument types is the dynamic signature itself.
\end{abstract}



\maketitle

\section{Introduction}

Intraday market manipulation is a moving target for regulators, and evaluating
detectors against it is usually blocked by a missing ingredient: ground truth.
This work uses two enforcement actions that, unusually, provide it. 

In the Indian BANKNIFTY index-options case,
aggressive trading in index constituents was timed to move the index in
favor of large options positions.
The footprint lasted minutes and spread across
thousands of option quotes. An interim order of the Securities and
Exchange Board of India (SEBI) supplies dated manipulation days,
enumerated with per-day strategy attribution in the order
itself~\cite{sebi2025interim}.

In \emph{SEC v.\ Patel} (2026), brought by the U.S. Securities and
Exchange Commission, a trader pumped
thinly traded U.S. equities with rapid-fire orders and dumped his
position within minutes. 
The complaint alleges more than a thousand such episodes.
The complaint's appendix names the tickers and the dates. 
Two markets, two
instruments, one dynamic signature: a fast, forced rise followed by an
immediate collapse.

A detector for this setting must satisfy three demands at once. 
(1) It must achieve high recall: a missed manipulation day is a regulatory failure.
(2) It must achieve workable precision: a system that flags a large fraction of normal days is ignored in practice.  
(3) It must be explainable: a regulator cannot act on an alert that merely says a day looks unusual. 
Regulators need to know which days, and why.

Preliminary experiments with our initial anomaly-detection baseline
exposed a central limitation: the model recovered every
regulator-identified day, but its precision was only about 30\%.
Closer inspection showed that it was responding to generic outliers
rather than the manipulation mechanism. Hourly aggregation can obscure
the signal further by erasing the minute-scale pattern before any model sees it
(Section~\ref{sec:signature}).

This work makes five contributions:
\bgroup\setlength{\topsep}{2pt}\setlength{\partopsep}{0pt}
\begin{enumerate}
\item A minute-level, strictly time-partitioned detection pipeline built
on smoothed state velocity. Its core feature is Delta-velocity,
motivated by the pump-and-crash signature of the manipulation.
\item An out-of-sample evaluation on a sequestered test period, with
thresholds fixed in advance. The plain autoencoder detects
\textbf{10 of 10} held-out regulator-identified manipulation days,
traced date-for-date to the enforcement order~\cite{sebi2025interim}.
\item A regime-aware analysis with a negative result. 
Volatility regimes 
inferred by a hidden Markov model (HMM) and selected by the Bayesian 
information criterion (BIC) 
are descriptively distinct, yet conditioning on 
them trades recall for precision.
The precision ceiling is consistent
with label incompleteness and was not resolved by any tested
architecture.
\item Cross-market, cross-instrument validation on enforcement labels.
In \emph{SEC v.\ Patel} (2026), the shape of the signature survives
transfer to thinly traded equities while velocity magnitude does not,
with the complaint's per-ticker, per-date record as ground truth.
\item Exact SHAP (SHapley Additive exPlanations) attribution over every
alert day (2{,}624 exactly explained rows, additivity error
$<10^{-18}$). Unconfirmed alerts share the regulator-identified
days' attribution profile. This is consistent with label
incompleteness and gives regulators an actionable explanation per
alert.
\end{enumerate}\egroup

Taken together, these contributions show that enforcement-linked ground
truth can anchor an explainable surveillance screen. Across the two
cases, what transfers is the minute-scale shape of market-state change,
not its raw magnitude.

\section{Related Work}

\subsection{Market Manipulation and Surveillance}

Market manipulation detection differs from ordinary anomaly detection because
the output is not merely a statistical warning; it must support regulatory
review. The economics of trade-based manipulation are documented
in~\cite{aggarwal2006stock}; the practical requirements of market
surveillance are framed in~\cite{aitken2015trade}, and contextual
time-series anomaly detection for manipulation appears
in~\cite{golmohammadi2015time}. Ensemble methods over hidden Markov
models support sequence classification in financial
data~\cite{kawawa2024ensemble}, and a patented stream-anomaly method
in this line was deployed at JPMorgan to flag inappropriate trading
behavior~\cite{balch2026method}; these systems operate on proprietary
behavioral data rather than public enforcement labels. These settings
emphasize a central difficulty:
manipulative trading is intentionally brief, strategic, and designed to resemble
ordinary market activity.

A further challenge is evaluation. Many manipulation-detection studies rely on
synthetic injections, simulated trading patterns, or self-labeled anomalies
because enforcement-linked manipulation labels are scarce; a recent ICAIF line of work
trains on synthetic data precisely because real labels are
rare~\cite{mahrous2025tstr}. A smaller thread uses enforcement material
directly, at coarse granularity: hourly blocks hand-labeled from 2003 SEC
lawsuits~\cite{diaz2011analysis}, later revisited with isolation-forest
ensembles~\cite{nunez2024ensemble}, and a small daily ticker-day benchmark
built partly from SEC documents~\cite{neela2025aimm}. This work extends that
lineage to minute-resolution, two-jurisdiction enforcement labels with
per-day strategy attribution. 
This makes our setting closer to forensic 
surveillance: we evaluate against enforcement-linked dates from documented
cases and treat unlisted days as unverified comparison days rather than true
negatives. This label structure motivates both our recall-first evaluation
and our caution in interpreting precision.

\subsection{Anomaly Detection in Financial Time Series}

Isolation Forest~\cite{liu2008isolation}, one-class
classifiers~\cite{scholkopf2001estimating}, and autoencoder-based models are
standard tools for unsupervised anomaly detection in financial time series,
with deep methods surveyed in~\cite{pang2021deep}. Their strength is that
they require little or no labeled manipulation data. Their weakness is that
they usually learn what is statistically rare, not what is manipulative.
The pump-and-dump literature illustrates both the promise and the label
problem: rule-based threshold detectors~\cite{kamps2018moon} and supervised
models over social-media-announced events~\cite{lamorgia2023doge} detect
crypto pumps whose labels come from the promoters' own channels, and
path-signature features, an encoding of trajectory shape, support
unsupervised detection in the same setting~\cite{akyildirim2022signature}.
These are the nearest precedents for shape-based scoring; none evaluates
against regulatory enforcement labels or options-market state.

This distinction matters in markets. Legitimate crashes, expiry effects,
liquidity shocks, and news events can all look anomalous. A detector that
scores each quote or time step independently may flag volatile days
without identifying the manipulation mechanism. Our approach keeps these
baselines, but changes the target feature: instead of detecting unusual option
levels alone, we detect unusual state movement through smoothed Delta velocity
and pump-reversal shape.

\subsection{Options Microstructure and Greek-Based Features}

Options manipulation differs from equity-price anomaly detection because the
relevant state is not only the traded price of the option or the underlying
asset. Option prices depend on moneyness, time to expiry, implied volatility,
and Greeks such as Delta and Gamma~\cite{black1973pricing,merton1973theory,hull2018options}.
A price move that appears large in raw option space may be ordinary once
strike, expiry, and volatility are accounted for. Conversely, a rapid
change in Delta can expose a short-lived dislocation that price levels
mute.

This structure motivates Greek-based features. Delta summarizes the option's
directional exposure to the underlying, while implied volatility captures the
market's option-specific risk assessment. 
Near expiry, Gamma effects can make Delta move quickly even without manipulation, which complicates detection.
Prior work on option market quality and liquidity also shows that option prices
and spreads vary systematically with contract characteristics and trading
conditions~\cite{mayhew2002competition,muravyev2020options}.
Our pipeline treats Delta velocity as a dynamic state variable
rather than a static Greek,
and evaluates whether the resulting signal separates manipulation-like
pump-reversal behavior from ordinary expiry and volatility effects.

\subsection{Explainable and Context-Aware Detection}

For surveillance, detection alone is insufficient: an alert must also be
interpretable and placed in a market context. 
Model-agnostic attribution methods
such as SHAP~\cite{lundberg2017unified} and LIME~\cite{ribeiro2016why} are
widely used to explain black-box predictions. 
In finance, explainability has been applied most often to fraud detection
in credit-card, accounting, and transactional data; options-market manipulation 
is less studied,
especially when the explanatory features are option-specific quantities such 
as implied volatility, Delta, and Delta velocity.

Context is equally important because an observation that is anomalous in a 
calm market may be ordinary during a volatile regime. 
Latent regime models, including hidden Markov models, trace back to classical 
regime-switching work in~\cite{hamilton1989new,rabiner1989tutorial} and have been 
applied to financial data in~\cite{nystrup2018dynamic}. 
We use regime inference not for forecasting
or portfolio allocation, but as a diagnostic test of whether regime-conditional
thresholds can separate manipulation-like velocity spikes from legitimate
volatility.

\subsection{Summary: The Gap This Work Fills}

Each of these threads solves part of the problem, and each stops short.
Surveillance studies name the conduct but rarely provide
enforcement-linked labels suitable for evaluation. Generic anomaly detectors find
outliers, not mechanisms. Option-pricing theory supplies the Greeks, but
treats them as static risk measures rather than detection features.
Attribution methods explain models, yet they are seldom applied to market
manipulation with instrument-specific features. Regime models describe
volatility well, but their value as a manipulation filter has not been
tested. 
This work brings all five together in a mechanism-specific
detector built on option-state dynamics. It is evaluated against
enforcement-linked dates in two markets, explained alert by alert, and
used to test the regime hypothesis. We are not aware of prior work that
combines these elements in an options-market manipulation setting.

\section{The Manipulation Signature}
\label{sec:signature}

The gap identified above starts with the signature itself, so we begin
there.

\begin{figure}
\centering
\includegraphics[width=\linewidth]{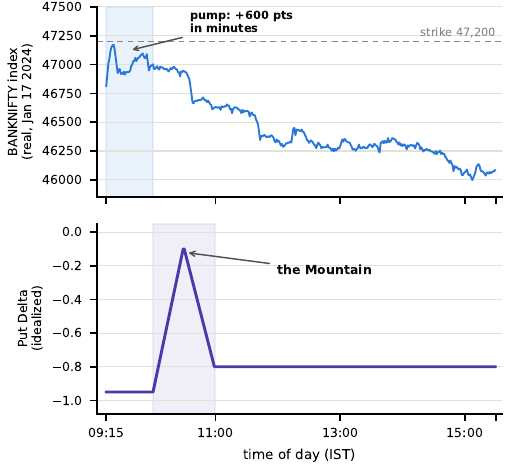}
\Description{Two stacked panels: real one-minute BANKNIFTY index on
January 17, 2024 with an early pump toward the 47,200 strike and a
day-long decline, above the idealized Put-Delta Mountain shape.}
\caption{The signature, real and idealized. Top: one-minute BANKNIFTY
index on January 17, 2024 (project data): a 600-point pump in minutes
toward the pinned 47,200 strike, then a day-long reversal. Bottom: the
idealized Mountain in Put-Delta space that the detector's velocity
feature targets. The event is dynamic; its price \emph{level} never
leaves the plausible range.}
\label{fig:mountain}
\end{figure}

\paragraph{The ``Mountain.''} The pump-and-reversal strategy produces a characteristic
pump-and-crash pattern. The option Delta is forced up sharply and then
crashes back, forming a transient ``mountain'' shape
(Figure~\ref{fig:mountain}) that decouples option behavior from the
underlying index. Near the strike, small index moves change Delta sharply.
This is a \emph{dynamic} event, not a high price
level.

\begin{figure}
\centering
\includegraphics[width=\linewidth]{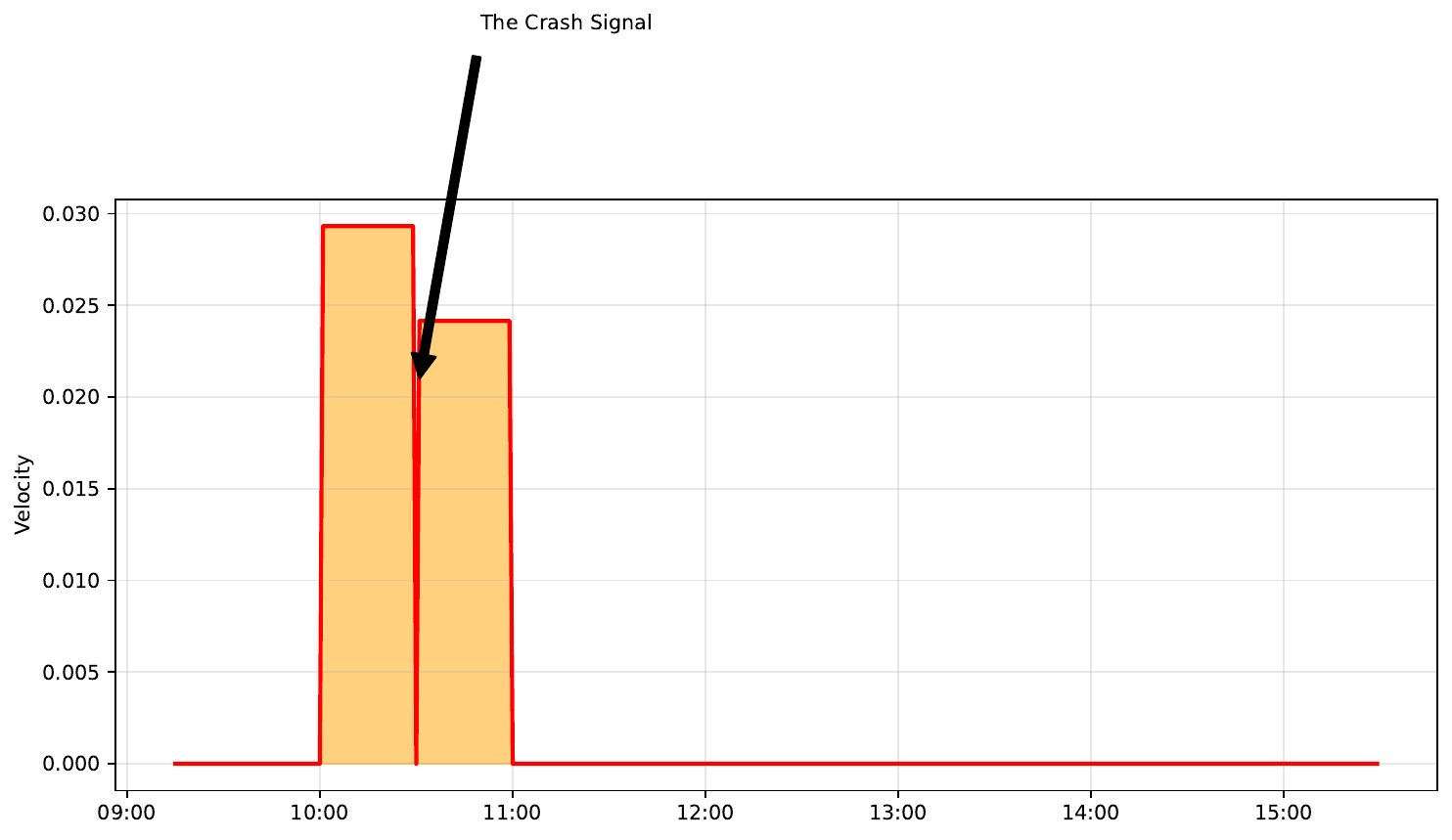}
\Description{Line chart of Delta velocity over time showing two sharp
spikes at the pump and the crash, flat elsewhere.}
\caption{The same event in velocity space: the pump and the snap-back
produce two spikes absent from normal trading: the feature the
detector consumes.}
\label{fig:velocity}
\end{figure}

\paragraph{From shape to feature: velocity.} Models cannot see shapes,
but they can see rate of change. The mountain creates two large velocity
spikes, one on the pump and one on the snap-back, that are absent in
normal trading (Figure~\ref{fig:velocity}). We engineer
Delta-velocity. For each day, strike, option type, and expiry, let
$\bar{\Delta}^{(15)}_t$ denote the trailing mean of 15 available Delta
observations and $\tau_t$ the timestamp of observation $t$. The
implemented feature is
\begin{equation}
v_t = \left|
\frac{\bar{\Delta}^{(15)}_t-\bar{\Delta}^{(15)}_{t-1}}
{(\tau_t-\tau_{t-1})/\mathrm{minute}}
\right|.
\end{equation}
On uninterrupted one-minute quotes, the smoothing span is 15 minutes.
Values are blanked when adjacent observations are less than 0.5 or more
than 2 minutes apart, and grouping by day prevents overnight differences.

\paragraph{Why levels and hourly bars fail.} Hourly averaging crushes a
10-minute spike into a small bump before any model sees it. Static Delta
levels capture position but not speed. Row-independent scoring breaks
the pump$\rightarrow$peak$\rightarrow$crash sequence. Velocity features
address all three limitations: minute-level resolution preserves the spike,
and the rate of change writes the local dynamics into every row, so even
a row-level model sees the sequence. Sequence architectures themselves
were tested and did not help (Section~\ref{sec:banknifty}).

\paragraph{Not every manipulation is a Mountain.} The enforcement record
attributes three of the eighteen days to a second strategy, extended
marking of the close: sustained selling pressure through the final
trading hour, aimed at the settlement price, with no pump and no
reversal~\cite{sebi2025interim}. That strategy paints no Mountain; on
its quietest day the index path is nearly flat. The distinction matters
for what follows: a signature detector should excel where the signature
exists and struggle where it does not, and Section~\ref{sec:banknifty} shows the errors
fall exactly along this line.

\section{Method}
The method has four parts. A minute-level feature captures the
signature. A time-partitioned protocol keeps the evaluation faithful to
the timeline. Regime conditioning tests whether market context separates
manipulation from volatility. Attribution explains every alert.

\subsection{Minute-level pipeline and Delta-velocity}\label{sec:pipeline}
We load BANKNIFTY spot and options data and resample them at one-minute
frequency. Delta is smoothed with a trailing mean of 15 available
observations within each (day, strike, type, expiry) group; on uninterrupted
one-minute quotes this spans 15 minutes. Velocity is divided by elapsed
minutes and blanked when adjacent observations are outside the 0.5--2.0-minute
interval. Grouping by day prevents overnight differences. The baseline
feature set is \{Option\_Price, Implied\_Vol, Delta\_Velocity\}. We
validated the Delta computation independently on 2024 U.S. options for
five liquid symbols, using one-minute quotes and vendor Greeks from
ThetaData. Across 56.5 million quote comparisons, the same
Black-Scholes computation reproduces vendor Delta at a Pearson
correlation of 0.9995 or better. Only isolated same-day-expiry put
strata exceed the locked error gates.

\subsection{Time-partitioned split and calibration}
The feature is only as credible as the protocol around it. Models are
trained on 2022 data, which contain no regulator-identified days.
Thresholds are calibrated on 2023 development data, and 2024 is held
out as a sequestered test period. We select thresholds on validation
with a recall-first row-quantile grid and \emph{lock them before}
examining any test label; train, validation, and test dates are
disjoint. Row-level decisions are aggregated into a daily anomaly hit
rate. This rate is more stable than a raw count because the number of
option rows varies by day.

\subsection{Regime-aware detection}
The central failure mode of a single global detector is that manipulation days
and legitimately volatile days occupy the same high-velocity tail. We test
whether conditioning detection on the prevailing market regime separates them.
A Gaussian hidden Markov model~\cite{rabiner1989tutorial} is fit on four
training-year market-state features: signed return, absolute return,
15-observation rolling volatility, and intraday range. 
The Bayesian
information criterion (BIC) 
selects four descriptively distinct states: 
state~0 is the calmest, state~3 the most volatile, and states 1--2 
are intermediate. 
Each observation
receives the state with the highest forward-filtered posterior
probability, so future returns do not determine its current state. 
We then build two regime-aware detectors. 
The \emph{hard regime} variant
uses a reconstruction model trained for the assigned regime, with a
global-model fallback for states that lack support. 
Its pooled scores
face one shared row threshold and one shared daily threshold,
calibrated on the validation year and locked before test scoring. The
\emph{posterior-weighted} variant instead blends reconstruction errors
using the HMM's state posteriors. The hypothesis is that
regime-specific norms stop flagging volatile but legitimate days while
manipulation remains extreme within its own regime.

\subsection{Explainable attribution}

Detection without explanation is not actionable for a regulator. For
each alert day, we compute exact SHAP values~\cite{lundberg2017unified}
on the plain autoencoder by enumerating all feature coalitions, then
aggregate row-level attributions into a per-day attribution profile.
SHAP explains how much each feature adds to or subtracts from a
baseline, the autoencoder's average reconstruction error over the
background sample, to reach a row's final error. 
These signed contributions are expressed in units of reconstruction error rather
than probabilities, and the three sum exactly to the row's deviation
from the baseline. 
Section~\ref{sec:explain} averages absolute values so contributions
of opposite sign do not cancel. The analysis asks two questions. First,
which features drive the model's alerts, and does the answer differ
between regulator-identified and unconfirmed days? Second, do the attribution
profiles of the highest-scoring \emph{unconfirmed} alerts resemble those
of regulator-identified days, consistent with undocumented events, or
do they resemble ordinary volatility, indicating a precision failure?
This turns the detector's mysteries into evidence a regulator can act
on.

\section{Experiments: BANKNIFTY}\label{sec:banknifty}
\paragraph{Setup.} The data are one-minute BANKNIFTY spot and options
bars for 2022--2024 from a public archival dataset~\cite{gupta2023banknifty},
with Greeks computed as in Section~\ref{sec:pipeline} and the
three-feature set \{Option\_Price, Implied\_Vol, Delta\_Velocity\}.
Models were trained on 2022 data, which contain no regulator-identified days.
Thresholds were calibrated on 2023, whose eight identified days served
as development ground truth, and were fixed before any 2024 label was
examined. The locked test year contains 209 trading days, of which ten
are regulator-identified manipulation days. All 18 dates match the
enforcement order's Table~44 exactly~\cite{sebi2025interim}; the order
attributes 15 of the 18 to an intraday pump-and-reversal strategy and 3
to an extended marking-the-close strategy (October~4, 2023 in the
calibration year; May~8 and July~10, 2024 in the test year), a
distinction Section~\ref{sec:discussion} returns to. Here, a positive label means a date
identified by the regulator as part of the alleged manipulation.
Unlisted dates are treated as negative for metric bookkeeping only;
they are unverified, not established as normal.

\begin{table}
\caption{Locked-test results, all seven models (2024: 10
regulator-identified days among 209). Precision, F1, and accuracy follow
the closed-world convention that every unlisted day is negative; the
labels do not verify that assumption, so reported precision is a floor.}
\label{tab:banknifty}
\small
\begin{tabular}{lcccc}
\toprule
Model & Recall & Alerts & Precision & F1\\
\midrule
Autoencoder & \textbf{10/10} & 41 & 24.4\% & \textbf{39.2\%}\\
HMM posterior-weighted AE & \textbf{10/10} & 41 & 24.4\% & \textbf{39.2\%}\\
One-Class SVM & 9/10 & 42 & 21.4\% & 34.6\%\\
Attention AE & 9/10 & 47 & 19.2\% & 31.6\%\\
Isolation Forest & 8/10 & 31 & 25.8\% & 39.0\%\\
Type-specific IF & 8/10 & 34 & 23.5\% & 36.4\%\\
HMM hard-regime AE & 7/10 & 26 & \textbf{26.9\%} & 38.9\%\\
\bottomrule
\end{tabular}
\end{table}

On the locked 2024 test set, the plain autoencoder detected \textbf{10
of 10} regulator-identified days (F1 = 39.22\%) with 41 alert days; the
Isolation Forest baseline detected 8 of 10 with 31 alerts, and the
one-class support vector machine (SVM) 9 of 10 with 42
(Table~\ref{tab:banknifty}). 
Under the closed-world convention, precision sits between 19\% and 27\% 
across 
seven models. 
This is the central tension the paper examines. 
No model dominates 
every objective: the plain and posterior-weighted autoencoders
maximize recall and F1, while the hard-regime 
model minimizes review volume at the cost of three misses.

\paragraph{Supplementary models.} The sequence-aware attention
autoencoder recovered 9 of 10 days but had the largest review load (47
alert days) and the lowest precision. A type-specific Isolation Forest
matched the baseline's 8 of 10. Added architectural capacity did not
add detection power on this feature set.

\begin{figure}
\centering
\includegraphics[width=\linewidth]{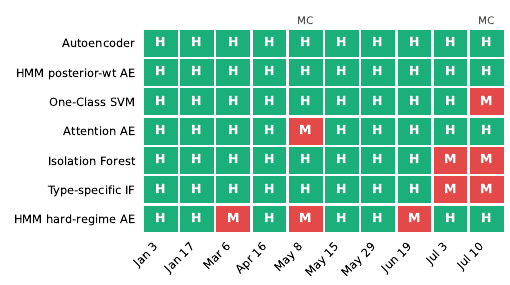}
\Description{Grid of seven models by ten test dates; nearly all cells
green (detected), with misses clustered on July 10, July 3, and May 8.}
\caption{Detection outcomes on the ten regulator-identified test dates.
Misses cluster on the two marking-the-close days (MC: May 8, July 10)
and July 3, though no single model misses both MC days; the plain and
posterior-weighted autoencoders detect all
ten.}\label{fig:matrix}
\end{figure}

\paragraph{The second strategy, seen in the errors.} Model errors are
structured rather than random: they follow the enforcement record's own
strategy attribution (Section~\ref{sec:signature}). The test year's two
marking-the-close days are where the weaker models most often fail.
Both Isolation Forest variants and the one-class SVM miss July~10; the
hard-regime and attention models miss May~8
(Figure~\ref{fig:matrix}, MC columns). 
These days have no Mountain to
see: their pressure is concentrated in the final hour and barely marks
the index path. Detection is marginal and footprint-dependent. Most
models catch May~8, whereas only the autoencoder variants recover
July~10.

A detector built for the pump-and-reversal signature reported,
accurately, which days contain that signature. The residual visibility
of the marking-the-close days suggests that pre-built positions and
final-hour activity still disturb the joint feature distribution. This
alignment is a strong tendency, not a rule. Both Isolation Forest
variants also miss one pump-and-reversal day, July~3, and the
hard-regime model misses March~6 and June~19. Because June~19 is the
highest-velocity day in the Section~\ref{sec:discussion} ranking, the latter misses
reflect regime routing rather than weak footprints.

\begin{figure}
\centering
\includegraphics[width=\linewidth]{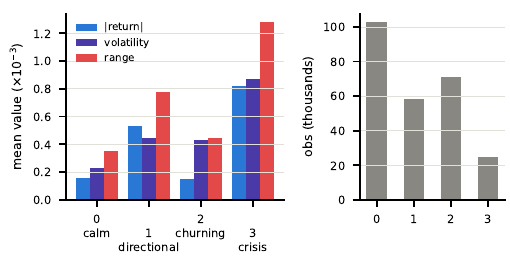}
\Description{Bar charts of the four HMM states' mean absolute return,
volatility, and range, plus observation counts per state.}
\caption{The BIC-selected four-state regime model, profiled over the
full 701-day period (256{,}677 minute rows). All three plotted features are
per-minute fractional moves of the index and share one dimensionless
scale: 1.0 on the axis is $10^{-3}$, or 0.1\% of the index level.
States are descriptively
distinct, from calm (state~0) to crisis (state~3); the intermediate
states do not order strictly by volatility.}\label{fig:hmm}
\end{figure}

\paragraph{Regime-aware results: a negative result.} The hard-regime
autoencoder reduces the alert load from 41 to 26 days and attains the
highest closed-world precision (26.9\%), but it detects only 7 of 10
regulator-identified days. 
The reduction in review volume costs recall. 
The posterior-weighted variant reproduces the plain autoencoder's daily
decisions exactly, with zero differences across the 454 validation and
test days that survive preprocessing.
Latent volatility regimes are BIC-selected and descriptively distinct
(Figure~\ref{fig:hmm}; state~0
the calmest, state~3 the most volatile), but regime awareness alone does
not resolve the manipulation--volatility overlap. A direct comparison
reaches the same conclusion. A regime-percentile variant and a long
short-term memory (LSTM) autoencoder each recover 8 of 10 days, with
31 alert days for the regime model and 39 for the LSTM. Regime
information buys review-burden efficiency, not recall.
The remaining bottleneck appears to be label scope, 
not model architecture.

\section{Experiments: Cross-Market Validation (SEC v. Patel)}\label{sec:patel}
\begin{figure}
\centering
\includegraphics[width=\linewidth]{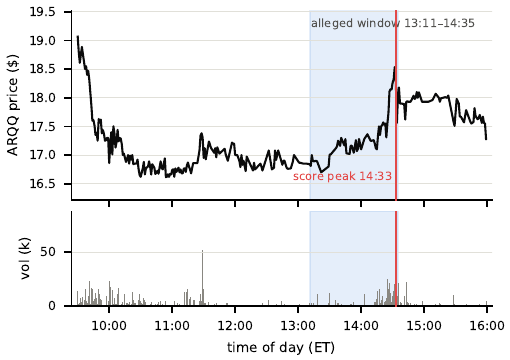}
\Description{ARQQ one-minute price and volume on October 8, 2021, with
the complaint's alleged window shaded and the model's score peak marked
at 14:33, at the top of the intraday rise.}
\caption{ARQQ, October 8, 2021 (one-minute bars). The complaint's
alleged window is shaded; the shape score's global peak (14:33) falls
inside it, at the start of the dump. ARQQ is thinly traded and does not
trade in every minute; the price line connects consecutive traded
minutes, and the volume panel shows where trading occurred.}\label{fig:arqq}
\end{figure}
We next test whether the dynamic signature transfers from Indian index
options to U.S. equities. In \emph{SEC v.\ Patel}~\cite{sec2026patel},
the complaint's Attachment~A names the manipulated tickers and dates,
providing rare per-event ground truth for U.S. markets. We focus first
on the complaint's two worked examples, ACY and ARQQ (35 alleged
ticker-days), using one-minute bars and a prespecified pump-reversal
\emph{shape} score: price rises over ten minutes and gives back within
five, weighted by volume participation.

The option-state features that feed the BANKNIFTY autoencoder (option
price, implied volatility, and Delta-velocity) do not exist for
equities. The shape score is the trajectory form of the same dynamic,
applied to the instrument that remains. Velocity magnitude alone
performs near chance in this setting (area under the receiver
operating characteristic curve, or AUC, near 0.5; a perfect ranking is
1.0). These securities are thinly traded, extreme minutes are common,
and the manipulator's footprint hides in that background. The shape
score, by contrast, ranks alleged days at AUC 0.91 [0.84--0.97] for
ARQQ (11 of 17 in the top-17) and 0.81 [0.73--0.87] for ACY, with
bootstrap confidence intervals from 10{,}000 resamples.

On the ARQQ worked example (October 8, 2021), the score's global peak
falls at 14:33, inside the complaint's alleged 13:11--14:35 window
(Figure~\ref{fig:arqq}). The complaint identifies that minute as the
start of the dump. The ACY result is weaker: its global peak falls
outside the alleged window, and its spring-2021 meme-rally period
produces organically mountain-shaped days. Some of those days may
themselves be unlisted events. Attachment~A begins only at the
defendant's May~25 broker warning, while the complaint's own worked
example predates it. The labels demonstrably undercount in both
markets.

\paragraph{Robustness on the broader universe.} A supplementary
workflow extends beyond the two worked examples. It uses 22 causal
microstructure features over 493 alleged ticker-days from the
complaint's wider universe, with a locked test of 31 alleged days
against 49 comparison days. 
A global autoencoder attains recall 0.968
at precision 0.517 (F1 0.674). A regime-conditioned variant routes days
with a hidden Markov model over lagged market-state features and
recovers all 31 alleged days, but at a higher review burden (71 versus
57 comparison alerts per 100 control days). The same label caveat
applies: comparison days are not verified negatives. The signature is
not an artifact of two selected tickers, and the regime lesson repeats
across markets. Regime conditioning trades review burden against
sensitivity; it does not change what is detectable.

\section{Explainability}\label{sec:explain}
Having evaluated detection and cross-market transfer, we return to the
BANKNIFTY alerts and ask what drove them.

\paragraph{Exact attribution over every alert.} The plain autoencoder
was frozen as an immutable checkpoint and replayed without refitting;
reconstruction scores matched saved values to within $5\times10^{-13}$.
Because the model uses three features, exact SHAP computation is
tractable: all $2^3=8$ feature coalitions can be enumerated, so no
approximation algorithm is needed. Using a deterministic 64-row
empirical background, we explained 64 fixed rows from each alert day:
2{,}624 rows across all 41 alert days, with maximum additivity error
$4\times10^{-19}$. Exactness refers to the attribution computation; the
64 rows from each day are a deterministic sample, not the day's full
record.

\begin{figure*}[t]
\centering
\begin{minipage}[t]{0.49\textwidth}
\centering
\includegraphics[width=\linewidth]{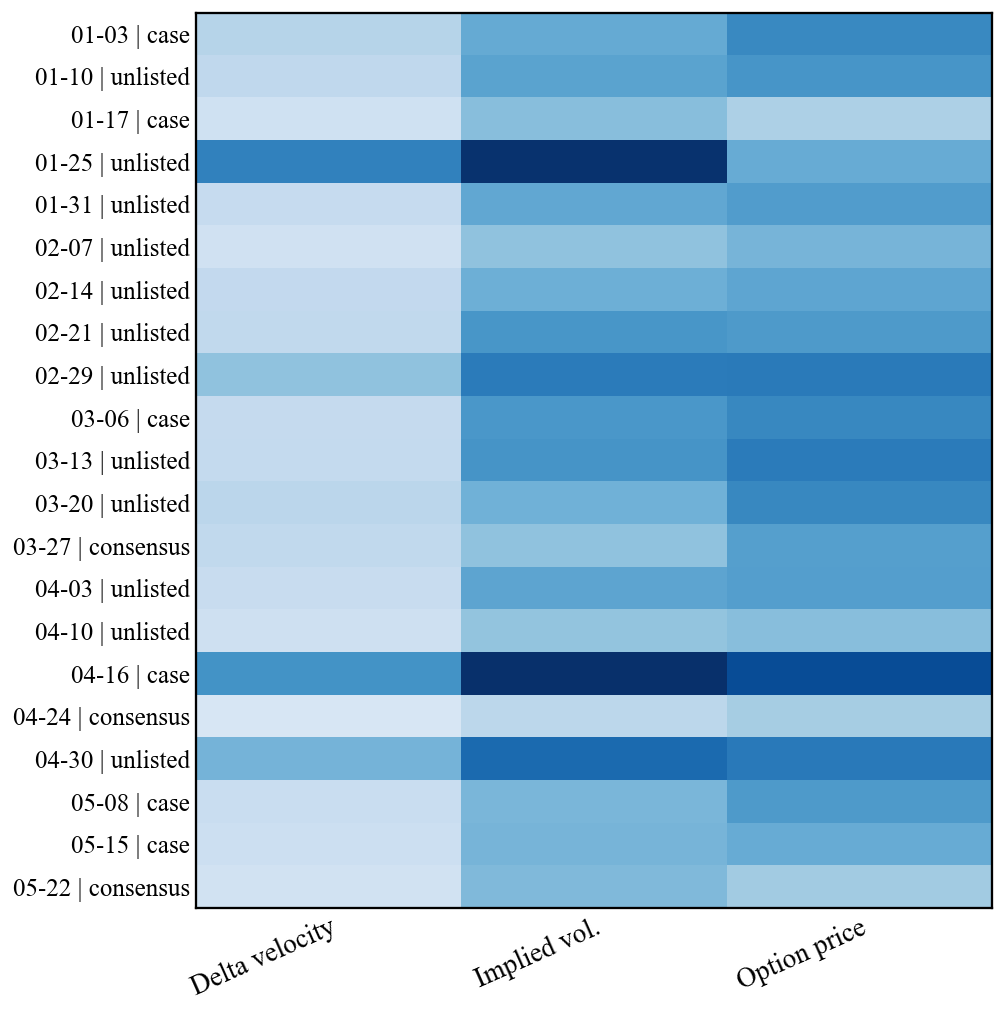}\\[-2pt]
\small (a) January--May 22, 2024
\end{minipage}\hfill
\begin{minipage}[t]{0.49\textwidth}
\centering
\includegraphics[width=\linewidth]{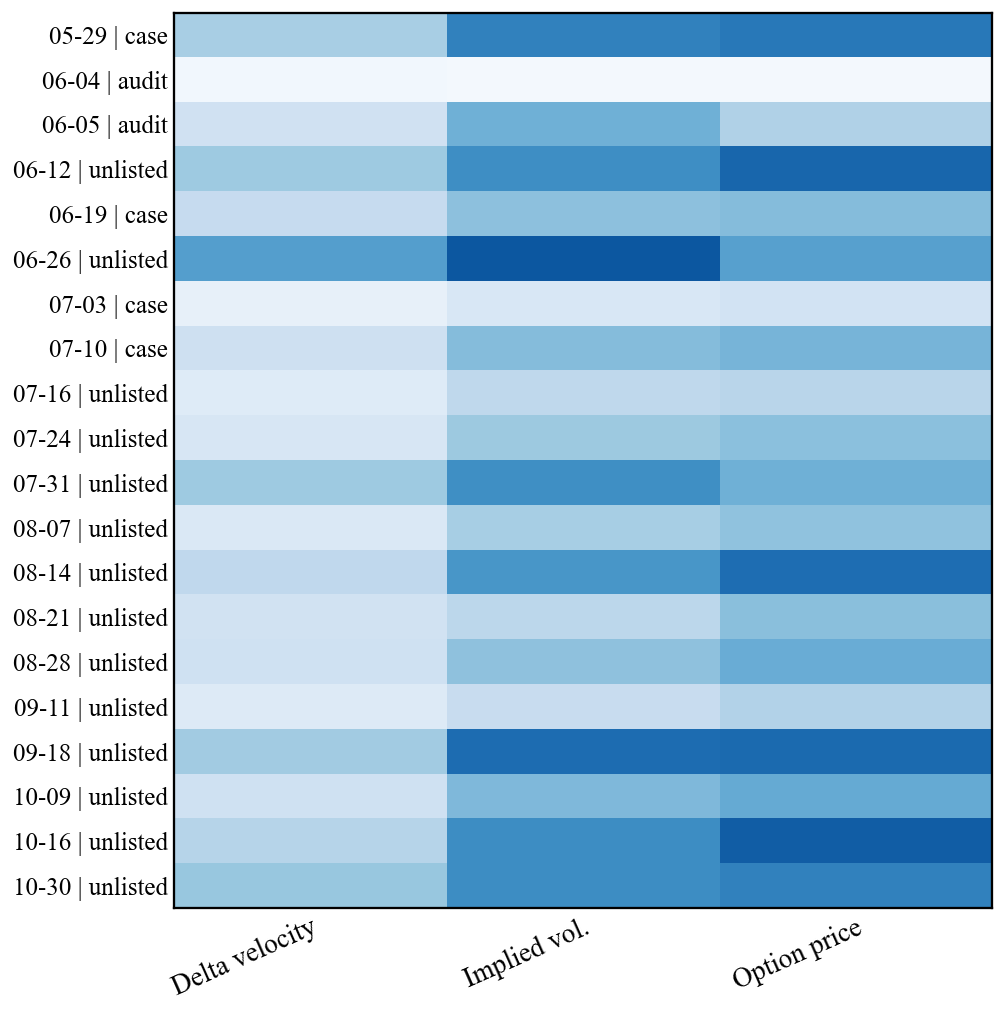}\\[-2pt]
\small (b) May 29--October 2024
\end{minipage}
\Description{Two heatmaps of per-day mean absolute SHAP attribution
magnitudes, with alert days as rows and the three input features as
columns. The shading pattern of regulator-identified rows resembles
that of unlisted, consensus, and audit rows.}
\caption{Per-day mean absolute SHAP attribution profiles for all 41
plain-autoencoder alert days of the locked 2024 test. Rows are alert
days, tagged \emph{case} (regulator-identified), \emph{consensus} (high
multi-model agreement), \emph{audit} (the June~4--5 election sessions),
or \emph{unlisted}; columns are the three input features. Both panels
share one color scale: color encodes mean absolute attribution
magnitude, and rows are not normalized to sum to one.
Regulator-identified and unconfirmed alerts share a common attribution
profile.}\label{fig:shap}
\end{figure*}

\paragraph{What drives the alerts.} Across all alerts, Option\_Price
carries 40.5\% of mean absolute attribution, Implied\_\allowbreak Vol 39.0\%, and
Delta\_\allowbreak Velocity 20.5\%, a pattern visible in
Figure~\ref{fig:shap} as the darker option-price and implied-volatility
columns. Delta-velocity is the feature that makes the
signature \emph{visible} at minute resolution (Section~\ref{sec:signature}), but it is not
the dominant attribution: the model alerts on a joint disturbance of
price, volatility, and dynamics. These shares are observed results, not
prespecified criteria, and they motivate framing the signature as a
\emph{shared attribution profile} rather than a single-feature tell.

\paragraph{Unconfirmed alerts share the profile.} The mean cosine
similarity between each unconfirmed alert's attribution profile and the
mean regulator-identified-day profile is 0.9894
(Figure~\ref{fig:shap}). 
Cosine similarity deliberately ignores
magnitude and tests whether alerts share the same attribution mix. In
this nonnegative attribution setting, it ranges from 0.0 (no overlap in
the contributing features) to 1.0 (identical attribution proportions).
The observed value of 0.9894 shows very strong consistency
between unconfirmed alerts and regulator-identified days. This supports
label incompleteness without relabeling any date as manipulation.
Because the comparison uses only three nonnegative features, we do not
treat it as a unique manipulation fingerprint; that is not its role
here, and a shared profile could also characterize extreme days in
general rather than manipulation in particular. 
A matched comparison points in the same direction. We paired the ten
identified days with ten unlisted non-alert days matched on realized
volatility, expiry distance, and weekday, yielding 494{,}626 rows. The
identified days still separated in Delta-velocity distribution, so the
contrast is not an artifact of cherry-picked calm days.

\paragraph{The regulator's reading.} For each alert the system reports
which features drove it and how closely its profile matches the
documented cases. 
The two highest-scoring unconfirmed alerts (June~4--5,
2024) coincide with the general-election result sessions and have two
plausible readings. They may reflect extreme legitimate volatility, or
they may contain activity resembling the documented mechanism during an
unusually volatile session. The system's output is a review priority,
not a verdict. High consensus on an unlisted day warrants contract-level
and event-context investigation before the alert is counted as either a
false positive or an undocumented event.

\section{Discussion and Limitations}\label{sec:discussion}
Precision remains the central challenge. Even with velocity features and
leakage-safe evaluation, single-threshold precision sits near 25\%, and
an LSTM autoencoder did not improve it:
manipulation and legitimate volatility are genuinely close in feature
space. Part of that closeness is adversarial. The enforcement record's
worked example was timed to a session that opened sharply lower on
earnings news, so the pump resembled dip-buying and the decline
resembled the day's own sentiment. 
By choosing already volatile
sessions as cover, a manipulator guarantees overlap with legitimate
volatility in any magnitude-based feature.

\paragraph{What suppression hides.} A ranking of all 455 trading days
in 2023--2024 by smoothed Delta-velocity makes the strategy split
concrete (rank~1 is the largest signal). The two weakest signals among
the eighteen identified days are extended-marking-the-close days:
July~10 ranks 115th and October~4 ranks 113th. The third
marking-the-close day, May~8, ranks 55th, consistent with its broader
detectability (Section~\ref{sec:banknifty}). By contrast, the pump-and-reversal day
June~19 ranks first across the full sample. The strategy associated
with most detector misses produces the least velocity signal.

The marking-the-close result also exposes a boundary of public data.
On July~10 the index's final hour is nearly flat. Only regulators'
participant-attributed data reveal the documented aggression: more than
35\% of market-wide constituent-futures volume in that hour. 
Public bars show observed price moves, not moves that were prevented. 
The two sources are complementary: a public-data screen identifies days that
merit attributed-data forensics, while the attributed layer is
indispensable for suppression-style conduct.

\paragraph{Label scope.} As defined in Section~\ref{sec:banknifty}, the labels are
positive-only: the order identifies alleged manipulation dates but does
not verify unlisted dates as normal. The order formed the labeled set by
first selecting the 30 most profitable days for minute-level review. A
day with the same pattern but smaller profit could therefore be absent
from the set. Reported
precision is a floor because the closed-world calculation counts every
unverified alert as a failure. This assumption can bias precision
downward; it does not change recall on the identified dates. Because the
source is a contested interim order, the dates are regulator-identified,
not adjudicated.

\paragraph{Magnitude is not economics.} Velocity magnitude tracks the
manipulation's profitability only weakly. Across the eighteen
identified days, the rank correlation between per-day smoothed velocity
and the order's per-day profit figures is 0.33 and not statistically
significant. The most profitable day, January 17, 2024, ranks only 28th
of 455 trading days by velocity. How much profit the strategy earned
and how strongly it moved Delta are different quantities. This is the
options-market form of the equities finding (Section~\ref{sec:patel}): the signature
is the shape of the move, not its size.

\paragraph{Scope and discipline.} All reported results come from a
single checkpointed run. An earlier development run held the feature
set and time split fixed but differed in architecture, feature
cleaning, training caps, cached rows, and calibrated thresholds.
Individual models' recall moved in both directions between the two
runs, so cross-run comparisons are not controlled experiments and are
not made here.
The threshold is an operational choice rather than a scientific one. A
label-free reference (the 85th percentile of validation-normal hit
rates) is close to the recall-calibrated value; in deployment, the cut
would reflect review capacity over ranked days. The Patel evaluation
covers the complaint's two worked-example tickers. It validates the
signature's transfer across markets and instrument types, not
universal generalization. On ACY, the unweighted shape score
outperforms the prespecified primary (0.87 versus 0.81), so both are
reported.
\section{Future Work}
These limitations motivate graph-based detection of the spot--option
decoupling as a cross-market edge anomaly; contrastive representation learning
(e.g., TS-TCC) to better separate legitimate volatility from manipulation;
streaming detection for near-real-time alerts; and a false-positive triage
workflow to surface the cause of each unconfirmed alert.

\section{Conclusion}
Two countries, two markets, two instrument types, one signature. A
regulator in Mumbai documented eighteen dated days of index-options
manipulation; a complaint in New York enumerated more than a thousand
pump-and-dump episodes in thinly traded equities. 
Records like these
are what detector evaluation has lacked. To our knowledge, this work is
the first to operationalize them at minute resolution in a leakage-safe
evaluation against enforcement-grade ground truth. With these records,
the pipeline recovered every held-out regulator-identified day in the
options case, ranked alleged days at AUC 0.91 in the equities case, and
attached an exact attribution to every alert it raised.

The enforcement order traced the same underlying dynamic by hand:
minute-wise parsing, index movement and Delta-exposure build-up
charts across thirty candidate days, assembled after the fact.
Converging on the same dynamic from opposite directions is strong
evidence it is the right thing to measure. This pipeline measures it
for every trading day and explains every alert; what was a forensic
reconstruction becomes a surveillance screen.

The pump-and-reversal manipulation, in both cases, was detectable 
not by how extreme the
market's state became but by the shape of how it moved. The
pump-and-reversal dynamic survives transfer across countries, markets,
and instrument types. Raw magnitude does not.

\bibliographystyle{ACM-Reference-Format}
\bibliography{refs}
\end{document}